\documentclass{article}

\usepackage{graphicx}   
\usepackage[margin=25mm]{geometry}

\usepackage[utf8]{inputenc}
\usepackage[T1]{fontenc}

\usepackage{listings}   
\usepackage{xcolor}     
\usepackage{amsmath}    
\usepackage{bm}         
\usepackage{siunitx}    
\usepackage{authblk}    
\usepackage{url}        
\usepackage{float}      

\newcommand{\abs}[1]{\left|#1\right|}
\newcommand{\avg}[1]{\left\langle #1 \right\rangle}
\newcommand{\fid}[1]{\avg{\abs{\gamma_#1}^2}}
\newcommand{\fidn}[1]{\abs{\gamma_#1}^2}

\newcommand{\fieldmod}{M_a}     
\newcommand{\fieldref}{R}       
\newcommand{\intensityoffset}{I_{\text{offset}}} 
\newcommand{\Ipa}{I^{(p)}_{ba}}

\begin{document}

\title{Practical considerations for high-fidelity wavefront shaping experiments} 
\author[1,2,3]{Bahareh Mastiani}
\author[1,2,4]{Daniël W. S. Cox}
\author[1]{Ivo M. Vellekoop}
\affil[1]{\footnotesize Biomedical Photonic Imaging Group, Faculty of Science and Technology, University of Twente, P.O. Box 217, 7500 AE Enschede, The Netherlands}
\affil[2]{These authors contributed equally to this work.}
\affil[3]{e-mail: mastiani.bahareh@gmail.com}
\affil[4]{e-mail: danielcox16@gmail.com}

\date{February 2024}

\maketitle
\hrule

\begin{abstract}
Wavefront shaping is a technique for directing light through turbid media. The theoretical aspects of wavefront shaping are well understood, and under near-ideal experimental conditions, accurate predictions for the expected signal enhancement can be given. In practice, however, there are many experimental factors that negatively affect the outcome of the experiment. Here, we present a comprehensive overview of these experimental factors, including the effect of sample scattering properties, noise, and response of the spatial light modulator. We present simple means to identify experimental imperfections and to minimize their negative effect on the outcome of the experiment. This paper is accompanied by Python code for automatically quantifying experimental problems using the OpenWFS framework for running and simulating wavefront shaping experiments.
\end{abstract}

\section{Introduction}
The ability to focus light through scattering media is critical for a wide range of applications, including deep tissue microscopy. However, it is challenging to focus light deep inside opaque media due to inhomogeneities in the refractive index of the medium. These inhomogeneities cause aberrations and scattering that prevent light from forming a diffraction-limited focus, thereby reducing the imaging resolution and contrast \cite{kubby2013adaptive}.

\begin{figure}
    \centering
    \includegraphics[width=0.98\linewidth]{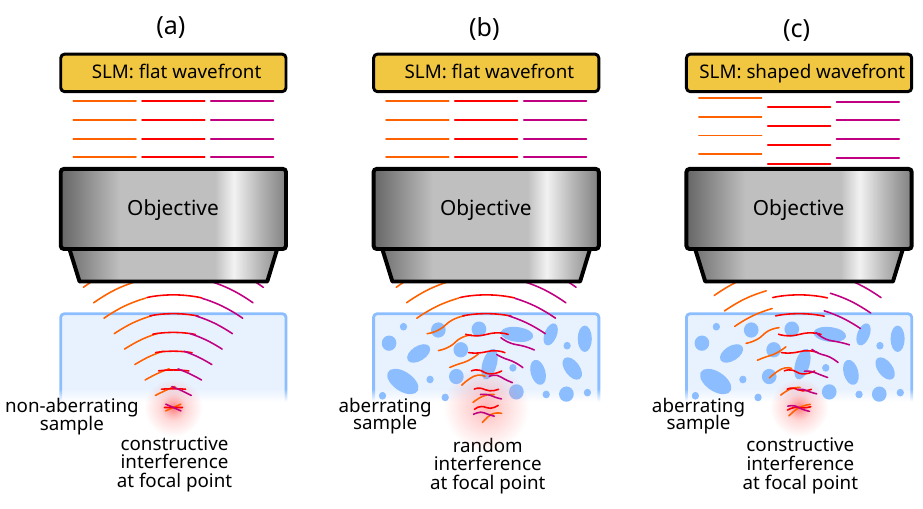}
    \caption{Schematic illustration of wavefront shaping. SLM: spatial light modulator. OBJ: objective. a. Focusing inside a non-aberrating sample. b. Focusing inside an aberrating sample results in a distorted focus. c. Focusing inside an aberrating sample with a WFS microscope. By shaping the incoming wavefront with a spatial light modulator, the focus can be restored.}
    \label{fig:wfs-concept}
\end{figure}

Wavefront shaping (WFS) is a powerful method for focusing light at target positions inside or behind scattering materials. The basic idea of WFS for microscopy is illustrated in Fig.~\ref{fig:wfs-concept}. Here, it is assumed that we use a scanning microscope, and we are interested in focusing the light inside the sample. In the ideal case (Fig.~\ref{fig:wfs-concept}a), the sample is transparent and the microscope objective focuses the light at the desired spot. We can describe this focusing as constructive interference, where each part of the converging wavefront interferes constructively at the focal point. When trying to focus light into an aberrating or scattering sample, (Fig.~\ref{fig:wfs-concept}b), light coming from different parts of the incident wave picks up different random phase deviations, causing random interference rather than constructive interference. As a result, the intensity in the focus is reduced, and the focus is spread out spatially. In the case of strong scattering, a disordered speckle pattern is formed. 

Wavefront shaping works by spatially modulating the incident wavefront to exactly compensate these random phase deviations, restoring constructive interference, causing the light to form a high-contrast, high-intensity focus inside the scattering sample (Fig.~\ref{fig:wfs-concept}c). Since the first demonstration of focusing light through opaque scattering objects \cite{vellekoop2007focusing}, the field has seen rapid development towards applications in fundamental research, microscopy, endoscopy, communication, and cryptography \cite{kubby2019wavefront, rotter2017light, gigan2022roadmap, cao2019reconfigurable, goorden2014quantum}.

The key goal in WFS is to find the unique incident wavefront that maximizes constructive interference in the target focus. A wide range of techniques are already available (see e.g. Refs.~\cite{kubby2019wavefront,horstmeyerGuidestarassistedWavefrontshapingMethods2015}), and exciting new approaches based on digital twins \cite{thendiyammalModelbasedWavefrontShaping2020, yamaguchiAdaptiveOpticalTwoPhoton2021a, Ploscher2015, Zhou2019} or deep learning \cite{gigan2022roadmap} are actively being developed.

One can easily compute the theoretical expected improvement of the focus quality under ideal circumstances \cite{vellekoop2007focusing}. In an actual experiment, however, the measured improvement may be significantly lower. Here, we address the following questions:

\begin{itemize}
    \item How much can I expect WFS to improve the focus?
    \item Why does my WFS experiment perform worse than expected?
    \item What can I do about it? 
\end{itemize}

This paper is supported by our Python package OpenWFS \cite{vellekoop2023OpenwfsGithub}, which includes code for \emph{automated analysis and troubleshooting} of WFS experiments. The package is available through the standard PyPI package repository \cite{vellekoop2023OpenwfsPyPI}, and documentation can be found at the Read the Docs website \cite{vellekoop2023OpenwfsReadthedocs}. Throughout this paper, we will refer to the relevant functions for detecting specific problems in WFS experiments.

For most of the analysis, we will assume that a `classical' WFS approach is used to find the optimized wavefront, that is, a phase-stepping technique that uses feedback from the desired focal point \cite{vellekoop2015feedback, popoff2010measuring, tao2017three, mastiani2021noise, mastiani2022wavefront}.
Although this covers only a fraction of the approaches in use, these classical algorithms have the advantage that they can be described analytically, allowing one to prove the convergence, analyze the effect of noise, and compute the expected improvement under a large range of circumstances. Moreover, many of the considerations here are generally valid regardless of the WFS algorithm used to find the optimal wavefront. 

After introducing the basic concepts that are needed to understand, 
run, 
quantify, 
and predict the outcome of a WFS experiment,
we cover the most common issues influencing the outcome.
We will explain how each of these issues affect the enhancement, how we detect or quantify them, and how to reduce their impact.
In addition, we will show that the effect of most of these issues can be quantified from regular phase-stepping measurements directly. This allows for an approach where troubleshooting metrics are conveniently built into the WFS algorithm at no extra cost. In addition, we provide a troubleshooter for detecting issues that do require specific measurement procedures to quantify them. We explain how to run the troubleshooter and analyze the troubleshooting metrics in section~\ref{sec:openwfs-troubleshooter}.

\section{Theory}
\label{sec:theory}
We can describe propagation of light through an arbitrary object with a matrix:
\begin{equation}
    \label{eq:tab}
    E_b = \sum_a t_{ba} E_a
\end{equation}
where $E_a$ is the incident field for mode $a$, and the summation runs over all optical modes. $E_b$ is the field at point $b$ where we want to focus the light. When focusing light through a scattering object, $t$ is the transmission matrix. Note, however, that this formalism is valid for any linear system, so $t_{ba}$ may also refer to propagation to a point inside a sample, or reflection, or in fact any linear system. Also note that the formalism does not depend on how we represent these modes (e.g. as a Hadamard or Fourier basis). We do restrict ourselves to orthonormal sets of modes, so that we have the simple expression
$P_\mathrm{in}=\sum_a\abs{E_a}^2$ for the the total incident power.

The aim of WFS is to find the incident field $E_a$ that maximizes the intensity\footnote{In this paper, we use the term `intensity' for the power in a single mode. When using a continuous representation of the transmission matrix (see Ref.~\cite{vellekoop2010exploiting}) $I_b\equiv \abs{E_b}^2$ in mode $b$, without changing the total incident power, $I$ is a true intensity with unit $W/m^2$, and all sums are replaced by surface integrals.}). The answer to this optimization problem is given by the Cauchy-Schwartz inequality, which states that:
\begin{equation}
    I_b = \abs{\sum_a t_{ba} E_a}^2 \leq \sum_a \abs{t_{ba}}^2 \sum_a \abs{E_a}^2\label{eq:Cauchy-Schwartz}
\end{equation}
The first term on the right hand side is a property of the sample. It can be understood in terms of a phase conjugation experiment: imagine we would place a point source in $b$, and observe the fraction of the total emitted power that propagates to all modes $a$ together. In this case, this fraction is exactly given by  $\sum_a \abs{t_{ba}}^2$, which we will denote $T_b$ from here forward. 

For a given sample and incident power, the right hand side of this inequality is fixed. It directly follows that at most a fraction $T_b$ of the incident light can be focused at $I_b$. This maximum is reached when both sides of Eq.~\eqref{eq:Cauchy-Schwartz} are equal, which is the case when $E_a\propto t^*_{ba}$, leading to an optimal value of
\begin{equation}
    I_b= \sum_a\abs{t_{ba}}^2 \sum_a\abs{E_{a}}^2 = T_b P_{in} \label{eq:I-b-optimal}
\end{equation}
All wavefront shaping algorithms for focusing light, in one way or the other, are aimed at determining $t_{ba}$ and subsequently displaying $t_{ba}^*$ on the SLM. 

\section{Experiment}
\subsection{Setup}
\label{sec:setup}
\begin{figure}
\begin{center}
	\includegraphics[width=0.80\linewidth]{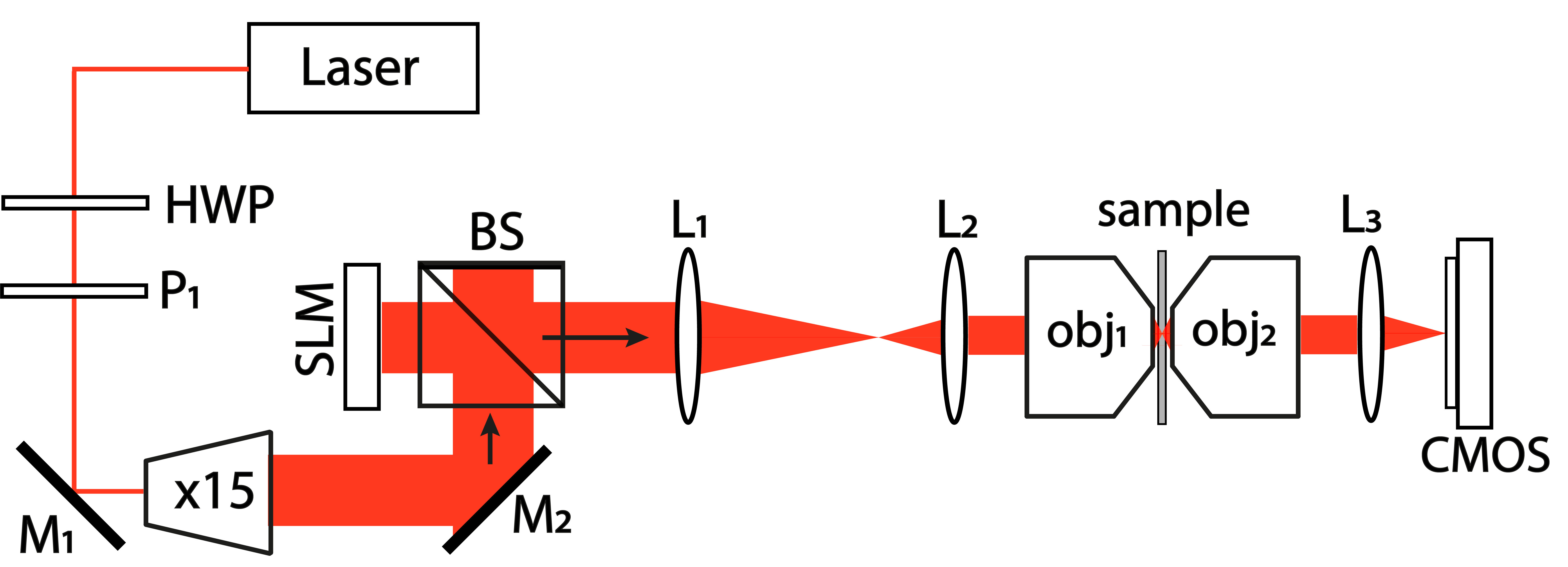}    
\end{center}
\caption{\label{fig:setup} Schematic of a simple WFS setup. HWP: half wave plate, P: polarizer, M: mirror, $\times$15: beam expander, SLM: spatial light modulator, BS: 50$\%$ non-polarizing beam splitter, obj: objective lens, $L_1: f = 200$ mm, $L_2: f = 75$ mm, $L_3: f = 75$ mm, $L_4: f = 150$ mm, $L_5: f = 50$ mm, CMOS: complementary metal oxide semiconductor camera.}
\end{figure}

A typical setup for WFS experiments is shown in Fig.~\ref{fig:setup}. In this setup, HeNe laser light (Thorlabs, \SI{2}{mW}, $\lambda$ = \SI{632.8}{nm}) is expanded by a beam expander (Thorlabs BE15M-A), and then modulated by a phase-only spatial light modulator (SLM, Hamamatsu LCoS-SLM, X13138 SLM) using a 50$\%$ beam splitter (Thorlabs CM1-BS013). A polariser ($P_1$, Thorlabs LPVISE100-A) is used to rotate the polarization to the linear polarization required by the SLM. A half wave plate (HWP, Thorlabs WPH10M-633) is used to control the total power entering the system. 

A 4f system images the SLM onto the back focal plane of a microscope objective (Zeiss A-Plan 100$\times$/0.8 M27), which focuses the light onto the sample. After transmitting through the sample, the light is collected by another objective lens (Zeiss A-Plan 100$\times$/0.8 M27), and recorded by a CMOS camera (Basler acA640-750um). A 4f system images the back surface of the sample onto a CMOS camera. This camera provides the feedback signal for WFS.

Note that there are many other options to provide the feedback signal. A mechanism to obtain a feedback signal that is correlated with localized light intensity is referred to as a `guidestar'. This guidestar can be a localized physical object (e.g. a fluorescent particle), as well as a localized physical phenomenon (e.g. multi-photon fluorescence or the photo-acoustic effect). \cite{horstmeyerGuidestarassistedWavefrontshapingMethods2015}

\subsection{Phase-stepping interferometry}
\label{sec:phase-stepping}
Each element of the transmission matrix $t_{ba}$ corresponds to an initially unknown complex optical field that must be measured. The phase of optical fields can typically not be measured directly. However, we can indirectly measure the relative phase of the field with a technique known as \emph{phase-stepping} or \emph{phase-shifting} interferometry \cite{LaiYatagai1991GeneralizedPhaseshifting}.

We will explain the use of phase-stepping interferometry with the \emph{Stepwise Sequential Algorithm} (SSA) as an example, but the technique is generally applicable. For SSA, the SLM pixels are divided into segments. We iterate over each segment $a$ to measure its contribution $\fieldmod = t_{ba}E_0$ to the field at feedback location $b$. $E_0$ is the field amplitude at the SLM surface, which is assumed to be equal for all segments. If the SLM is illuminated unevenly, we can include the illumination profile in the matrix element $t_{ba}$ (c.f. Section~\ref{sec:phase-only}). The rest of the segments are used as static reference and produce the field $\fieldref$ at $b$.

In each iteration, we phase-modulate the segment of interest by a set of phases $ \phi_p $ to produce the field $\fieldmod e^{i\phi_p}$ at $b$. In other words, the relative phase between the fields is varied between $0$ (including) and $2\pi$ (excluding) in $P\geq 3$ steps. Due to the interference between light originating from the modulated segment $\fieldmod e^{i\phi_p}$ and light originating from the other segments $\fieldref$, the feedback intensity will respond sinusoidally:

\begin{equation}
    \label{eq:phase-step-I_p}
	\Ipa = |\fieldmod e^{i\phi_p} + \fieldref|^2 = |\fieldmod|^2 + |\fieldref|^2 + \fieldmod^* \fieldref e^{-i\phi_p} + \fieldmod \fieldref^* e^{i\phi_p}
\end{equation}
where
\begin{equation}
    \label{eq:phase-step-phi_p}
	\phi_p = \frac{p}{P}2\pi,
	\qquad p = 1, ..., P
\end{equation}

$\Ipa$ corresponds to the intensity in mode $b$, measured when applying phase $\phi_p$ to the modulated input mode $a$. Note that the terms with $\fieldmod \fieldref^*$ and $\fieldmod^* \fieldref$ carry a factor $e^{\pm i \phi_p}$. By combining the measurements $\Ipa$ in a \emph{Discrete Fourier Transform} (DFT), these terms can be isolated:

\begin{equation}
    \label{eq:phase-step-dft}
	F_{k,a} \equiv \frac{1}{P}\sum_{p=1}^{P} \Ipa e^{-ik\phi_p},
\end{equation}
where $k$ is an integer in the range:
\begin{equation}
    \label{eq:phase-step-k}
    k =
    \begin{cases}
        -P/2, \:\ldots, \:P/2-1, &\text{ if $P$ is even} \\
        -(P-1)/2, \:\ldots, \:(P-1)/2, &\text{ if $P$ is odd}
    \end{cases}
\end{equation}

It follows that $F_{1,a} = \fieldmod \fieldref^* = t_{ba}E_0R^*\propto t_{ba}$. So we can find the optimal input wavefront as:
\begin{equation}
    \label{eq:phase-stepping-Ea-from-Fk}
    E_a^\text{opt} \propto \left(F_{1,a}\right)^*
\end{equation}
In theory $\Ipa$ follows a sinusoid with an offset, hence only $F_{-1,a}, F_{0,a}$ and $F_{1,a}$ are non-zero. However, in real measurements, the values for $\abs{k}>1$ are never exactly zero, due to noise in the measurements and imperfect SLM response calibration (see Section~\ref{sec:noise}).

\section{Enhancement}
One way to quantify the improvement of the focus, is the fraction of the incident power that is focused into mode $b$, i.e.,  the fraction $S\equiv I_b/P_{in}$. In adaptive optics, it is usually assumed that all light is scattered in the forward direction, so that $T_b=1$, and $S$ is called the \emph{Strehl ratio} \cite{kubby2013adaptive}. From Eq.~\eqref{eq:I-b-optimal}, it is direct that in this case the optimal wavefront achieves perfect diffraction-limited focusing: $S=1$.

For a scattering sample, it is often not possible to determine exactly what fraction of the incident power reached the focus. Therefore, it makes more sense to use a relative measure $\eta\equiv I_b/I_0$, which is called the \emph{enhancement} \cite{vellekoop2007focusing}. The reference intensity $I_0$ is technically defined as the intensity in point $b$, averaged over all possible `similar' samples for a given optimized incident wavefront \cite{vellekoop2015feedback}, i. e.
\begin{equation}
    I_0 = \avg{\abs{\sum_a \tilde{t}_{ba} E_a}^2},
\end{equation}
where $\tilde{t}_{ba}$ is the transmission matrix for a different sample with the same scattering strength, thickness, etc., and $\avg{\cdot}$ denotes averaging over all such samples.

For strongly scattering samples, $\tilde{t}_{ba}$ and $t_{ba}$ will be completely uncorrelated. Moreover, we assume that all elements $t_{ba}$ are statistically independent and identically distributed for all modes $a$ \cite{goodman2015statistical}, so that $\avg{t_{ba}^* t_{ba'}}=\delta_{a a'}\tau$  where $\delta_{aa'}$ is the Kronecker delta, and the value $\tau=\avg{\abs{t_{ba}}^2}$ is an average `mode-to-mode' transmission coefficient which does not depend on mode index $a$. We can now use these statistical properties and insert $E_a=E_0 t_{ba}^*$  into Eq.~\eqref{eq:I0} to find

\begin{align}
    I_0 &= E_0^2 \avg{\sum_{a,a'} \tilde{t}^*_{ba} \tilde{t}_{ba'} t^*_{ba}t_{ba'}}\\
    &=E_0^2 \sum_{a,a'}\delta_{aa'}\tau\; t^*_{ba}t_{ba'}\\
    &= E_0^2 \tau \sum_a \abs{t_{ba}}^2 = \tau P_\text{in}\label{eq:I0}
    \end{align}
In practice, $I_0$ can be determined by translating the sample, while keeping the shaped wavefront constant, and averaging measurements taken at different positions. Alternatively, the average speckle background can often be used \cite{vellekoop2015feedback}. Expressions for the expected enhancement as a function of the number of modes, and for phase-only light modulation are given in Section~\ref{sec:modes} and Section~\ref{sec:phase-only}, respectively.


\section{Fidelity}
In an experimental setting, the generated wavefront always deviates from the optimal one, so that the fraction of light that reaches the focus is always lower than $T_b$. To account for these imperfections, we describe the incident field as a superposition of the optimal field and a residual field that is orthogonal to the optimal one \cite{vellekoop2008universal}:
\begin{equation}
E_a=\gamma E_a^\text{opt} + \sqrt{1-\abs{\gamma}^2} E_a^\text{res}.
\end{equation}
Here, the incident fields are normalized such that $\sum_a \abs{E_a^\text{opt}}^2 = \sum_a \abs{E_a^\text{res}}^2$.
The factor $\gamma$ denotes the fraction of the incident field that is shaped correctly. It is defined as the correlation coefficient of the ideal wavefront $E_a^\text{opt}$ and the actual wavefront $E_a$ \cite{vellekoop2008universal}:

\begin{equation}
\gamma \equiv \frac{\sum_a \left(E_a^\text{opt}\right)^* E_a}{\sqrt{\sum_a\abs{E_a}^2}\sqrt{\sum_a\abs{E_a^\text{opt}}^2}}.\label{eq:fidelity-definition}
\end{equation}
By definition, the residual field does not contribute to the target intensity at all. Therefore, the fraction of light that reaches the focus equals $\abs{\gamma}^2 T_b$. The factor $\abs{\gamma}^2$ is called the \emph{fidelity}, and equals the fraction of the incident power that is shaped correctly. For ideal control over the incident wavefront $\abs{\gamma}^2=1$. In practice, however, there are many effects that reduce the fidelity. In this paper, we will treat the following effects:
\begin{itemize}
    \item The effect of a finite number of modes $\fidn{N}$
    \item The effect of phase-only modulation $\fidn{A}$
    \item The effect of non-modulated light $\fidn{0}$
    \item The effect of imperfect phase modulation $\fidn{\phi}$
    \item The effect of phase jitter $\fidn{J}$
    \item The effect of a finite signal-to-noise ratio $\fidn{S}$
    \item The effect of drift and sample decorrelation $\fidn{D}$
\end{itemize}
For each of these values, we will estimate the \emph{expected} value, averaged over noise and over all `statistically similar' samples; that is, samples that are macroscopically the same, but differ in the microscopic placement of the scattering particles. This averaging is denoted by $\avg{\cdot}$.

We assume that each of these effects is uncorrelated with the other, meaning that we can multiply the fidelities and simply estimate the expected intensity in the target focus as
\begin{equation}
\avg{I_b}=\fid{D}\fid{0}\fid{\phi}\fid{J}\fid{S}\avg{\fidn{A}\fidn{N}T_b} P_{in}.\label{eq:fidelity-stack}
\end{equation}
Here, all factors were assumed to act independently, except for the terms $\fidn{A}\fidn{N}T_b$, which were averaged together, as described in Sections~\ref{sec:modes} and \ref{sec:phase-only}.
Clearly, it is essential that all fidelities are as close to $1$ as possible. For each of these factors, we now describe how to estimate it from the experimental data, and what measures can be taken to increase it if it is too far below $1$.

\section{Limited control}
\subsection{Finite number of modes\label{sec:modes}}
So far, we considered $\sum_a$ a summation over \emph{all} optical modes. In practice however, a WFS experiment only controls a fraction of the incident angles and positions, and often only a single polarization. We now analyze how this limitation affects WFS fidelity. 

First, consider the case of \emph{ideal phase and amplitude modulation}, where the SLM generates the optimal field $E_a = E_0 t^*_{ba}$ for a finite number of modes $a\leq N$, and that $E_a=0$ for $a>N$. Inserting this field into Eq.~\eqref{eq:fidelity-definition} gives 
\begin{equation}
\fidn{N} = \frac{\left[\sum_a^N E_0\abs{t_{ba}}^2\right]^2}{ \sum_a^N E_0^2 \abs{t_{ba}}^2 \sum_a \abs{t_{ba}}^2} = \frac{\sum^N_a\abs{t_{ba}}^2}{\sum_a\abs{t_{ba}}^2}.\label{eq:fidelity-limited}
\end{equation}

Unsurprisingly, this ratio equals $1$ when all optical modes are considered, and it reduces when $N$ decreases. Intuitively, one can interpret $\fidn{N}T_b=\sum_a^N\abs{t_{ba}}^2$ as the fraction of the light that would propagate from a light source at point $b$ into the $N$ modes controlled by the SLM in a `time-reversal' experiment.

For each additional mode, we need additional measurements to determine $t_{ba}$. When fewer modes are considered, the WFS process will be faster, at the cost of reducing the fidelity. In general, not all modes contribute equally. In this case, it is beneficial to choose a set of modes that contributes most to the intensity, and discard all other modes. For example, for a sample that mainly scatters light in the forward direction, one can choose a basis consisting of smooth modes that correspond to small angular deviations \cite{mastiani2022wavefront}. For a homogeneously illuminated SLM, we can thus use the same expression as for ideal phase and amplitude modulation (Eq.~\eqref{eq:fidelity-limited}). 

A similar approach is taken in adaptive optics, where a set of low-order Zernike modes may be used \cite{kubby2013adaptive}. There is, however, a key difference: in adaptive optics the modes are used to express the phase of the light rather than the field. This results in a highly non-linear response for phase variations $\SI{>1}{rad}$, causing many of the algorithms used in adaptive optics to fail if aberrations are strong \cite{siemonsRobustAdaptiveOptics2021}.


\begin{figure}
\begin{center}
	\includegraphics[width=0.6\linewidth]{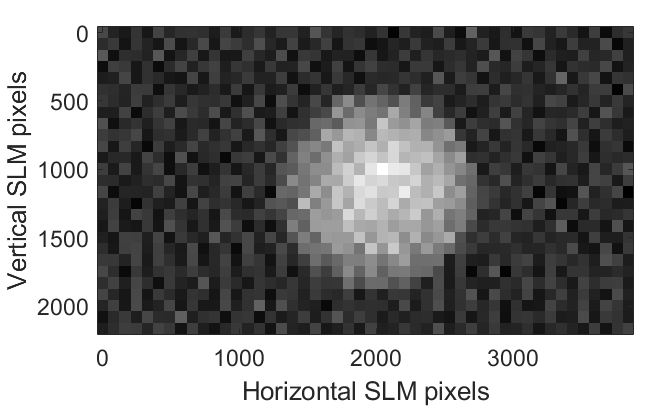}    
\end{center}
\caption{Measured beam location on the SLM. Each segment's displayed brightness corresponds to the measured contribution $\abs{t_{ba}}^2$ for that segment.} \label{fig:SLM illumination} 
\end{figure}

In practice, the set of modes should be chosen correctly even for a strongly scattering sample. For example, in the pupil-conjugate geometry shown in Fig.~\ref{fig:setup}, the SLM is mapped to the back pupil of a microscope objective. If we choose modes that correspond to (groups of) pixels, some modes may correspond to pixels that are blocked by the back pupil of the objective (see Fig.~\ref{fig:SLM illumination}). Clearly, including these modes in the WFS process costs time and does not contribute to the enhancement of the focus.

\subsubsection{Enhancement in strongly scattering samples\label{sec:enhancement}}
For strongly scattering samples, we can use the assumptions described in Section~\ref{sec:theory} to compute the estimated enhancement. Firstly, we use the fidelity in Eq.~\eqref{eq:fidelity-limited} to find the optimized intensity of
\begin{equation}
I_b=\fidn{N}T_b P_\text{in}=\sum_a^N\abs{t_{ba}}^2 P_\text{in}.
\end{equation}
Together with the reference intensity $I_0=\tau P_\text{in}$ (Eq.~\eqref{eq:I0}),
we get an enhancement of 
\begin{equation}
    \eta\equiv\frac{I_b}{I_0}=\frac{\sum_a^N\abs{t_{ba}}^2}{\tau},\label{eq:enhancement}  
\end{equation}
Which averages to 
\begin{equation}
    \avg{\eta}=N.
\end{equation}

Therefore, for strongly scattering samples and ideal WFS, the expected enhancement is equal to the number of segments. Due to the experimental factors discussed below, in practice a lower value (around 0.5 N for strongly scattering samples in our typical experiments) results.

\subsubsection{Automatic detection and quantification}
\label{subsubsec:slm-illumination-detect}
After running a WFS algorithm, we have access to the measured values of $\abs{t_{ba}}^2$ for each mode $a$ (see Section~\ref{sec:phase-stepping}), which we call the \emph{contribution} of that mode. By plotting the contribution for each mode, as in Fig.~\ref{fig:SLM illumination}, we can verify if all modes contribute approximately equally, or whether some modes may be omitted. In practice, the modes that contribute will often have a different contribution depending on their position on the SLM or shape of the generated optical mode. In Section.~\ref{sec:phase-only}, we explain how to quantify the effect of this distribution automatically.

\subsubsection{Mitigation}
If the center and the size of the SLM pattern do not match the beam center and size observed in the above measurement, these parameters must be manually corrected. In OpenWFS the pattern can be translated and scaled with the \texttt{transform} argument of the SLM object. In addition, the algorithm may be written to skip modes that do not contribute strongly. Large gains can selecting only the modes that contribute most, such as the modes in a Fourier basis that correspond to small angle scattering \cite{mastiani2022wavefront}.

When modes correspond to physical (groups of) pixels on a phase-only SLM, care must be taken not to choose the pattern too small. If there is a significant contribution of non-modulated light (from pixels outside the pattern), this will also negatively affect the fidelity. See Section~\ref{sec:non-modulated-light}.

\subsection{Phase only modulation\label{sec:phase-only}}
Often, a phase-only SLM is used in experiments. As a result, it is not possible to match the amplitude distribution of the incident field to that of $t_{ba}$, resulting in a reduction of the fidelity. We now analyze how this effect reduces the fidelity beyond the reduction caused by selecting a finite number of modes, $\fidn{N}$ . For ideal phase-only light modulation, we have an incident field $E_a=E_0 t_{ba}^*/\abs{t_{ba}}$, in a basis where indices $a$ correspond to pixels on the SLM. Using the definition of the fidelity (Eq.~\eqref{eq:fidelity-definition}) again, we find
\begin{equation}
\fidn{A} = \frac{\left[\sum_a^N E_0 t_{ba}\; E_0 t^*_{ba}/\abs{t_{ba}}\right]^2}{\left(\sum_a E_0^2\right)\left(\sum_a^N E_0^2\abs{t_{ba}}^2\right)} = \frac{\left[\sum^N_a\abs{t_{ba}}\right]^2}{N\sum_a^N \abs{t_{ba}}^2}=\frac{\overline{\abs{t_{ba}}}^2}{\overline{\abs{t_{ba}}^2}},\label{eq:fidelity-phase}
\end{equation}
where $\overline{\phantom{x}}$ denotes the averaging over all pixels of the spatial light modulator. 

Note that we assumed here that the SLM is illuminated homogeneously. However, the illuminated beam on the modulator usually has Gaussian beam profile resulting in a lower fidelity. We can simply include this illumination profile in $t_{ba}$ so that $E_0$ is a constant, and we can still use Eq.~\eqref{eq:fidelity-phase}.

\subsubsection{Enhancement in strongly scattering samples}
If all modes contribute equally on average, we can multiply the enhancement $\eta$ in Eq.~\eqref{eq:enhancement} by the fidelity in Eq.~\eqref{eq:fidelity-phase} to find the enhancement for phase-only modulation
\begin{equation}
    \eta=\frac{\left[\sum_a^N\abs{t_{ba}}\right]^2}{N \tau}.\label{eq:enhancement-phase-1} 
    \end{equation}
To compute the expected value for $\eta$ we use the fact that the elements $t_{ba}$ for a strongly scattering sample follow a circular Gaussian distribution \cite{goodman2015statistical}. This distribution has the property that $\avg{\abs{t_{ba}}}^2=\frac{\pi}{4}\avg{\abs{t_{ba}}^2}=\frac{\pi}{4}\tau$.
Expanding the Eq.~\eqref{eq:enhancement-phase} , and ensemble averaging, we find
\begin{align}\eta&=
\frac{1}{N\tau} \avg{\left[\sum_a^N\abs{t_{ba}}\right]^2}\\
&=\frac{1}{N\tau}\left(\avg{\sum_{a,a'\neq a}^N\abs{t_{ba}}^2} + \avg{\sum_a^N\abs{t_{ba}}^2}\right)\\
&=\frac{1}{N\tau}\left(\frac{\pi}{4} N (N-1)\tau + N\tau\right)
\end{align}
giving the well known result \cite{vellekoop2007focusing}
\begin{equation}
    \avg{\eta} = \frac{\pi}{4}(N-1) + 1.\label{eq:enhancement-phase}
\end{equation}
Note that the conditions leading to this result are almost never fulfilled exactly. If the SLM is not illuminated evenly, or not all modes contribute equally, we have to fall back to Eq.~\eqref{eq:fidelity-phase}. For example, even if the sample is completely random and $\avg{\abs{t_{ba}}^2}$ does not depend on $a$, we have to take into account the illumination profile on the SLM, giving \cite{vanPutten}
\begin{equation}
    \fid{A}\equiv\frac{\overline{A_a}^2}{\overline{A_a^2}}, 
\end{equation}
with $A_a$ the field amplitude of the illumination at mode $a$.

Finally, note that for weak, forward scattering, we can approximate $t_{ba}=e^{i\phi_{ba}}$. This means that there is no need to modulate the intensity of the light and $\fidn{A}=1$.

\subsubsection{Automatic detection and quantification}
This fidelity $\fidn{A}$ in Eq.~\eqref{eq:fidelity-phase} is computed in OpenWFS as the \texttt{fidelity\_amplitude} attribute in the \texttt{WFSResult} object returned by a WFS algorithm. If this value drops below $\pi/4\approx 0.79$ significantly, this is an indication that not all modes contribute equally, which may be caused by an uneven illumination of the SLM.

\subsubsection{Mitigation}
\begin{figure}
\begin{center}
	\includegraphics[width=0.6\linewidth]{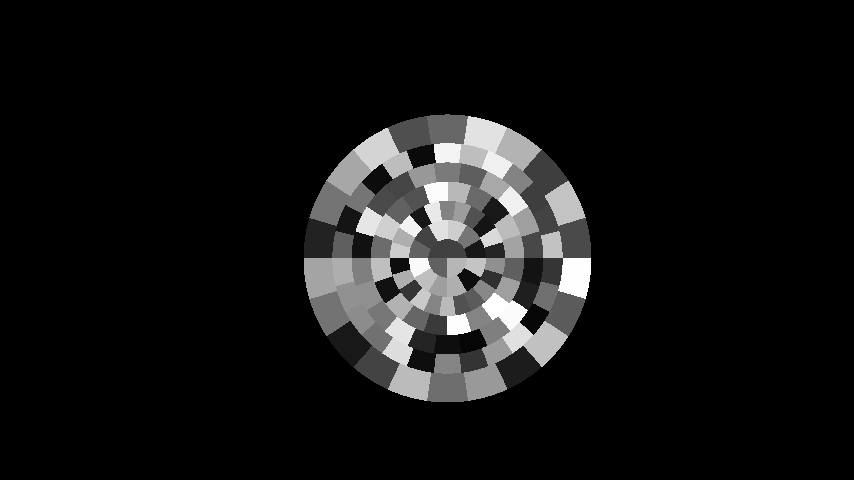}
\end{center}
\caption{\label{fig:SLM-circular-geometry} SLM screen with a random wavefront displayed on the circular geometry. Each segment's displayed brightness corresponds to the applied phase. The segment sizes are adjusted to compensate for the beam intensity profile.}
\end{figure}

Ref.~ \cite{mastiani2021noise} pioneered the use of a circular geometry to mitigate the effect of a non-uniform illumination of the SLM (see Fig.~\ref{fig:SLM-circular-geometry}). The idea is to adjust the segment size to the intensity of the illumination, using larger segments in regions where the intensity is lower. This way, we are able to have each segment, on average, contribute equally to the WFS procedure. For this geometry, the segment size can be changed by adjusting the number of segments per ring and the ring radii. In OpenWFS, this segment distribution is available as the function \texttt{slm.geometry.circular}.

\section{Imperfections of the spatial light modulator}
In this section, we describe the typical limitations of liquid crystal on silicon (LCoS) phase only SLMs, describe how these limitations affect the fidelity, and provide strategies to mitigate these effects.

\subsection{Non-modulated light\label{sec:non-modulated-light}}
\begin{figure}
\begin{center}
	\includegraphics[width=0.5\linewidth]{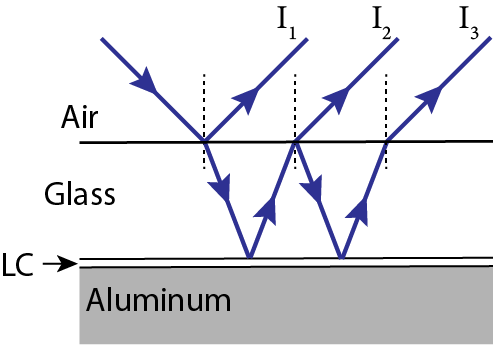}
\end{center}
\caption{\label{fig:SLM reflection} The main transmitted/reflected beams in the simplified SLM layer structure.}
\end{figure}

Most liquid-crystal SLM devices operate in reflection mode. As a result of the layer structure of the SLM there are two main reflections. Fig.~\ref{fig:SLM reflection} shows the simplified layer structure and the predominant transmitted/reflected beams.
The first reflection ($I_1$) is generated on the front surface of the protective glass. The main beam ($I_2$), is reflected at the back surface, thus traversing the liquid-crystal (LC) layer, and is the most intense reflection.
Finally, internal reflections inside the protective glass can result in dim higher order reflections ($I_3$) \cite{martinez2014analysis}.

The front surface reflection ($I_1$) of the SLM is not modulated and thus does not contribute to the shaped wave. The reflection coefficient at the air-glass interface depends on the SLM model and the wavelength of the incident beam.
For instance, the amplitude reflection coefficient at the air-glass interface of the SLM, model X10468-08 from Hamamatsu, is measured as 0.198 for $\lambda=\SI{647}{nm}$ \cite{martinez2014analysis}.

If we assume that the non-modulated light does not contribute to the intensity in the focus, we can find the fidelity
\begin{equation}
    \fid{0}=\frac{I_\text{modulated}}{I_\text{total}}.
\end{equation}
Note that for nearly transparent samples, the non-modulated light may still contribute to the focus, so that this equation is not generally valid. Finally, it should be noted that the popular parallel-aligned (PAL-LCoS) SLMs only modulate one of the polarization states. Any component of the light in the perpendicular polarization state contributes to non-modulated light. 

\subsubsection{Automatic detection and quantification}
OpenWFS provides a simple means to estimate the amount of non-modulated light originating from the front surface reflection of the SLM. It simply performs phase stepping with a flat wavefront. This results in an intensity of 
\begin{equation}
    I_b^{(p)} \propto |E_0\sqrt{1-\gamma_0^2} + E_0 \gamma_0 e^{i\phi}| + \intensityoffset,
\end{equation}
with $\gamma_0$ the fraction of the field that is not modulated, and $\intensityoffset$ the offset in the detector signal caused by non-coherent sources, such as amplifier bias or ambient light. We first determine $\intensityoffset$ by closing the shutter of the laser, and subsequently subtract this value from the signal. The amount of non-modulated light can now be estimated from the phase-stepping data as
\begin{equation}
    \frac{\abs{F_1}}{F_0}=\gamma_0 \sqrt{1-\gamma_0^2}.
\end{equation}
Assuming that $\gamma_0^2 > \frac12$, we find
\begin{equation}
    \gamma_0^2 = \frac{1+\sqrt{1-4 \frac{\abs{F_1}^2}{F_0^2}}}{2},
\end{equation}
which is output by the \texttt{troubleshoot} function of OpenWFS in the property \texttt{fidelity\_non\_modulated}. Note that this function produces incorrect results in the extreme case that more than half of the light is not modulated, such as when the SLM is accidentally turned off. 

\subsubsection{Mitigation}
High ratio's of non-modulated light can be caused by a pattern of SLM segments that underfills the illuminated SLM area, which can be solved by adjusting the mapping of segments to the SLM surface (see Section~\ref{subsubsec:slm-illumination-detect}).
 
A different source of non-modulated can be the previously explained front reflection (see Fig.~\ref{fig:SLM reflection}). The reflectivity of the surface depends on the model of the SLM, and on the wavelength-specific anti-reflection coating applied to it.  If the front reflection is too strong, it is possible to tilt the SLM, and aim the front reflection $I_1$ onto a beam dump. Of course this also introduces tilt in the main reflection $I_2$, but this tilt can conveniently be compensated for with a phase gradient on the SLM. Note that this technique is typically unsuitable for pulsed light. The steep phase gradient typically surpasses well beyond $2\pi$ and must therefore be phase-wrapped. Hence, the pulse is delayed by different amounts on different parts of the SLM due to phase wrapping. 

Lastly, non-modulated light may be caused by the light not having the correct polarization. PAL-SLMs only modulate one polarization component of the light. By choosing an appropriate state of the incident polarization, we can minimize the non-modulated light and thus optimize the fidelity.

\subsection{Imperfect SLM calibration}
\label{sec:slm-calibration}
In order to display an image on the SLM, the SLM hardware receives the image and converts pixel values to voltages with the use of an internal lookup table. Each SLM requires calibration to attain a linear phase response between 0 to 2$\pi$ for the wavelength of the incident laser light. Any deviation from this linearity introduces a phase error for each input mode, and reduces the fidelity of the WFS experiment. Therefore, it is crucial to calibrate the SLM properly and load the correct lookup table to the SLM hardware. Alternatively, OpenWFS provides the option to specify a software lookup table, which is accelerated by the GPU using OpenGL.

\begin{figure}
\begin{center}
	\includegraphics[width=0.5\linewidth]{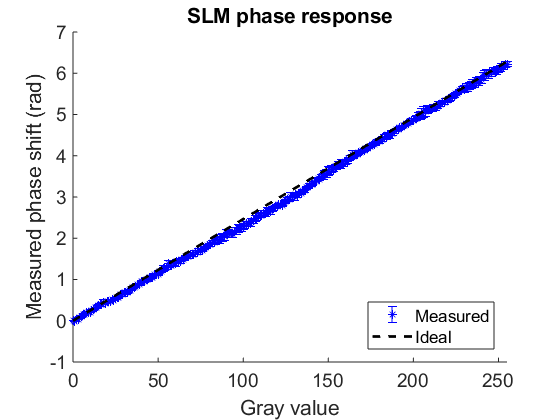}    
\end{center}
\caption{\label{fig:SLM phase response} Measured SLM phase response and the ideal phase response (dashed line) versus the gray values displayed on the SLM.}
\end{figure}

Figure~\ref{fig:SLM phase response} shows an example of the measured phase response for the SLM (Hamamatsu X13138-07) averaged over three trials. The measured phase response closely matches the ideal linear phase response (dashed line) between 0 to $2\pi$, which demonstrates that the proper lookup table is loaded on the SLM for the chosen wavelength of incident laser light. 

If the response of the SLM deviates from a linear phase response, the fidelity of displaying the wavefront reduces. To account for this effect, we assume that the light modulator only modulates the phase, which is appropriate for parallel-aligned (PAL) SLMs, but only approximately so for the older twisted-nematic (TN) SLMs.  Let $f(\phi_p)$ denote the actual phase as a function of the intended modulation phase $\phi_p$. The response of the light modulator can now be written as a Fourier series
\begin{equation}
    \label{eq:mod-factor-expif-as-fourier}
    e^{i f(\phi_p)} = \sum_k c_k e^{i k \phi_p},
\end{equation}
where the sum is over the range of $k$ as in Eq.~\eqref{eq:phase-step-k}. Since we assumed phase-only modulation, the coefficient $c_0=0$. Note that this coefficient, if present, would not effect the quantification measurements described below anyway, and the effect of this non-modulated light is already accounted for in $\fidn{0}$, as explained in Section~\ref{sec:non-modulated-light}. The $c_1$ component corresponds to the intended phase response, and all other components correspond to orthogonal modes that on average do not contribute to the focus. We can thus quantify this effect with the fidelity
\begin{equation}
    \fid{\phi}=\frac{\abs{c_1}^2}{\sum_k \abs{c_k}^2}.
\end{equation}
Note that for a perfect calibration $f(\phi) = \phi$, $c_1=1$ and all other $c_k=0$.
Also note that an imperfect phase calibration also reduces the signal-to-noise ratio (SNR) in a measurement, causing an additional drop in the fidelity (see Section~\ref{sec:noise}).

\subsubsection{Automatic detection and quantification}
The effect of an imperfect calibration on the phase-stepping response Eq.~\eqref{eq:phase-step-I_p} can be written as:
\begin{align}
    \label{eq:wrong-calibration-I_m}
	I^{pa} &= |\fieldmod e^{i f(\phi_p)} + \fieldref|^2
    = |\fieldmod|^2 + |\fieldref|^2 + \fieldmod \fieldref^* e^{i f(\phi_p)} + \fieldmod^* \fieldref e^{-i f(\phi_p)}\\
    &= |\fieldmod|^2 + |\fieldref|^2 + \fieldmod \fieldref^* \sum_k c_k e^{i k \phi_p} +  \fieldmod^* \fieldref \sum_k c_k^* e^{-i k \phi_p}
\end{align}
The response is no longer a single sinusoid. This fact emerges in the form of non-zero higher frequency components $F_{k,a}$ for $\abs{k}>1$. Taking the DFT as defined in Eq.~\eqref{eq:phase-step-dft} now yields:
\begin{align}
    \label{eq:wrong-calibration-Fk}
    F_{k,a} = \begin{cases}
        |\fieldmod|^2 + |\fieldref|^2 &\text{ for $k=0$} \\
        c_k \fieldmod \fieldref^* + c_{-k} \fieldmod^* \fieldref&\text{ otherwise}
        \end{cases}
\end{align}

For phase stepping, we only use $F_{1,a}$. Note that if non-modulated light is present, we can include it in the $R$ term and still choose $c_0=0$. Also note that the signal is affected by the $c_{-1}$ component of the phase response.  The effect of a non-zero $c_{-1}$ component cannot be separated from the measured response directly. Here, we  assume that $c_{-1}\ll c_1$, so that we can neglect it. In this case, $F_{1,a}\approx c_1 M_a R^*$ and 

\begin{align}
    F_{k,a} = \frac{c_k}{c_1}F_{1,a} + \frac{c_{-k}}{c_1^*} F^*_{1a}\qquad\text{ for $k\neq 0$}.\label{eq:F-with_calibration}
\end{align}
To isolate the $c_k$ components, we perform a linear-least-squares fit. This is done by constructing a matrix with two columns: $F_{1,a}$ and $F_{1,a}^*$, and computing the Moore-Penrose pseudoinverse of this matrix: $F^+$. The components of $c$ are then estimated from the measurements as
\begin{equation}
    \begin{bmatrix}
        c_k/c_1\\
        c_{-k}/c_1^*
    \end{bmatrix}= F^+
    \begin{bmatrix}
        F_{k,1}\\
        F_{k,2}\\
        \ldots\\
        F_{k,N}
    \end{bmatrix}\label{eq:pseudo-inverse}
\end{equation}
giving a fidelity
\begin{equation}
    \fid{\phi}=\frac{1}{\sum_{k\neq 0}{\abs{c_k/c_1}^2}},
\end{equation}
which equals $1$ for a perfectly calibrated SLM. Computing $\fid{\phi}\leq 1$ provides a way to check for systematic phase errors in the SLM response directly from the WFS measurements. Note the pseudoinverse optimally uses the data of all measurements, making it robust even in the presence of noise. Lastly, note that if $c_{-1}$ is not negligible, some of the other components $c_k$ need to be non-zero as well, in order to fulfill the phase-only constraint $\abs{e^{if(\phi)}}=1$. Therefore, we find a fidelity $<1$ in this case too, so we can still detect a problem with the SLM calibration. The only exception is when $c_{-1}=1$, meaning that the SLM modulates the phase in the exact opposite direction we expect. This case cannot be detected with this method. For feedback-based WFS this is not a problem at all, since the phase sign flips occur both in the measurements and the construction of the focus and thus cancel out. For model-based WFS, however, the sign of the phase modulation is important and should be checked independently\cite{thendiyammalModelbasedWavefrontShaping2020}.

The value of $\fid{\phi}$ is computed automatically from the WFS measurements if the number of phase steps is larger than three. It is available in the WFSResults structure as \texttt{fidelity\_calibration}.

\subsubsection{Mitigation}

\begin{figure}
\begin{center}
	\includegraphics[width=0.6\linewidth]{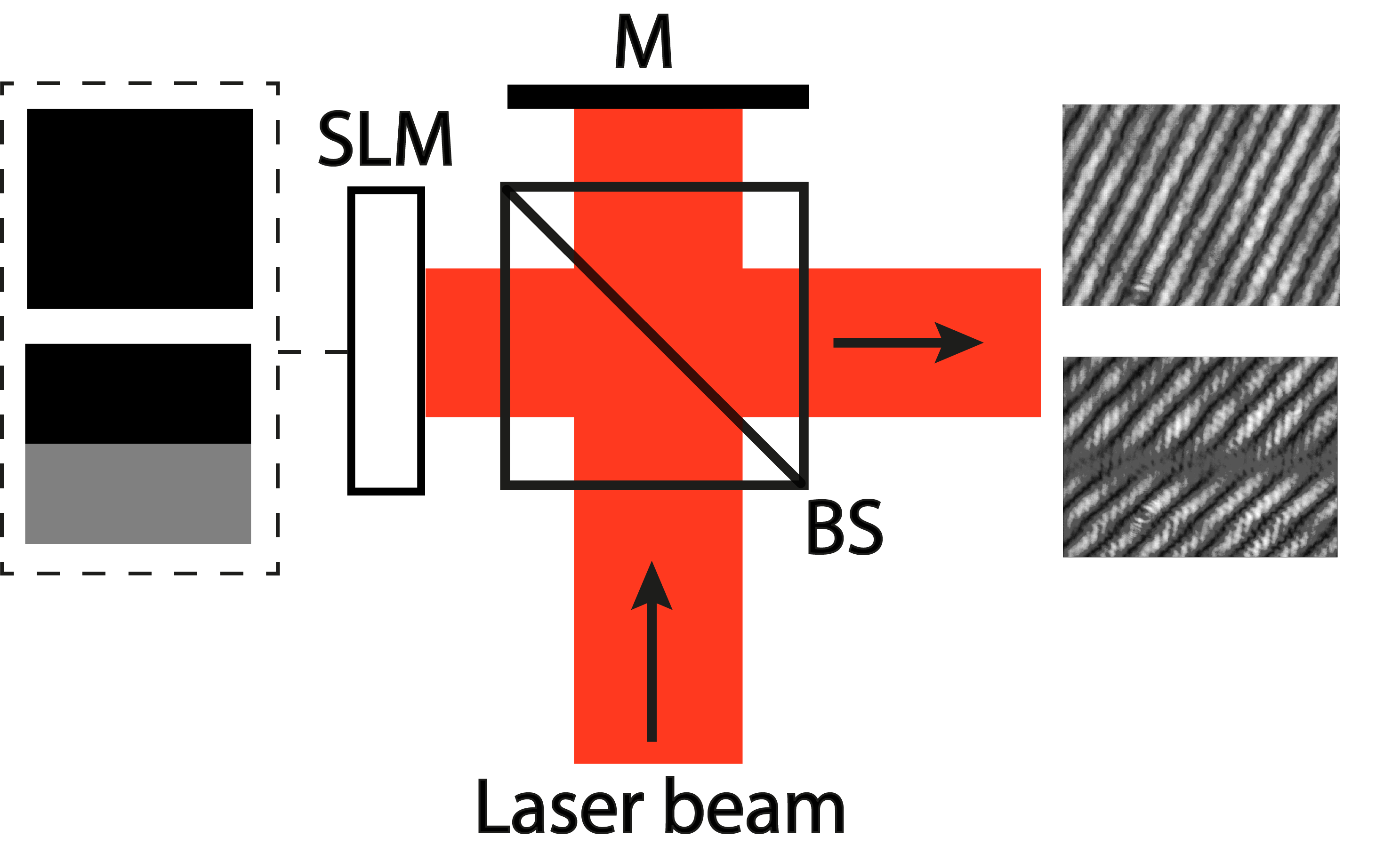}    
\end{center}
\caption{\label{fig:SLM calibration} Schematic of the setup for calibrating the SLM lookup table. M: mirror, BS: beam splitter.}
\end{figure}

A common method to calibrate the phase response of the SLM is using a Michelson-Morley interferometer. Figure~\ref{fig:SLM calibration} shows a simplified schematic of the calibration setup. The laser beam reflected from a mirror (M) interferes with the modulated beam reflected from the SLM. The SLM is divided horizontally in two halves. The phase of the top half (or the bottom half) is modulated from 0 to $2\pi$. This causes the fringe pattern in the top half to shift with respect to the fringes in the bottom half. Since both top and bottom halves of the beam follow the same path, this approach is insensitive to drift in the path length of one of the arms of the interferometer. To define the phase shift value in the fringes, we first perform Fourier transforms of the top and bottom quarters of the fringes image. Neglecting the offset value (first Fourier component), we find the Fourier component having the maximum amplitude which corresponds to the spatial frequency of the fringes. The phase difference of this maximum-amplitude component defines the phase of the fringes. Finally, the measured phase response is inverted and loaded into the SLM hardware as a lookup table. Lastly, note that several other calibration procedures exist. \cite{Li2019ProgressPhase}

\subsection{SLM phase jitter}
\label{sec:slm-phase-jitter}
For LCoS SLMs, the effective phase applied by each pixel is not perfectly stationary. Typically, the phase fluctuates periodically with a frequency equal to the SLM refresh rate or a multiple thereof. This effect is known as \emph{phase jitter} or \emph{flicker} \cite{Garcia-Marquez2012FlickerMinimization}, and it causes the displayed phase pattern to deviate periodically from the phase pattern that was sent to the SLM. For a uniformly distributed phase error between $-\Delta \phi$ and $\Delta \phi$, the fidelity reduction is given by \cite{chandrasekaran2014using}
\begin{equation}
\fid{J}=\left(\frac{\sin \Delta\phi}{\Delta \phi}\right)^2. \label{eq:fidelity_phase_error}
\end{equation}
It can be seen that small errors in the phase do not contribute significantly to the overall fidelity. However, this phase jitter also causes noise during the measurements, which is especially prominent if the algorithm needs to isolate a small variation on top of a large background of modulated light. The phase jitter also adds noise to the WFS measurements, causing a further reduction of $|\gamma|^2$ (see Section~\ref{sec:noise}).

\subsubsection{Automatic detection and quantification}
The SLM phase jitter can be distinguished from regular noise by physically turning off the SLM and observing the effect on the noise. The phase jitter may be quantified by performing the interference calibration explained in Section~\ref{sec:slm-calibration}, but instead of modulating the phase of one of the halves, both SLM halves remain constant and are observed at a high frame rate. Hence, for this method the required observation frame rate is several times higher than the refresh rate of the SLM.

\subsubsection{Mitigation}
The effect of phase jitter on the measurements can be reduced by setting the exposure time of a camera, or the integration time of a detector, to a multiple of the period of the jitter. This way, the jitter largely averages out over the measurement. However, this does not reduce the effect of jitter in constructing the wavefront, so a factor of $\fid{J}$ still arises there.

To reduce the amount of phase jitter, we may use the lowest possible voltages on the SLM that still cause a full $2\pi$ phase shift, and choose zero grey value for flat reference wavefronts whenever possible because the jitter is typically higher at higher voltages. It may also be higher for specific gray values due to the pulse code modulation scheme applied in many SLMs. The amount of jitter is also dependent on the model and vendor of the SLM, and can be further reduced by cooling the SLM \cite{Garcia-Marquez2012FlickerMinimization}.

\subsection{SLM timing}
An important aspect in optimizing SLM performance is correct timing of the measurements. If the measurement is started too early, the liquid crystals are not yet settled into their new state. If the measurement is started too late, it wastes photon budget and needlessly increases the duration of the experiment.

The image on the modulator is updated with the specified refresh rate. For some SLMs, a hardware clock output is available, or may be isolated from the video signal. OpenWFS, instead, automatically syncrhonizes the measurements using a software triggering mechanism, where OpenGL is used to wait for the \emph{vertical retrace} event, a historical name for the start of a new frame, and update the image during that event. It is essential to only update the SLM image during the vertical retrace: if it is updated before or after \emph{tearing} occurs, causing the SLM to display part of the old frame and part of the new frame simultaneously. 

After the vertical retrace, it takes some time for the image information to be transferred from the PC to the SLM hardware, and it may also take some additional time for the SLM hardware to start updating the voltages on the SLM pixels. During this \emph{latency}, or \emph{idle time}, there is no optical response of the SLM at all. We found that an idle time of 1-3 frame periods is typical, depending on the SLM hardware.

The second interval, the settle time, is the time that is needed for the liquid crystals of the SLM to stabilize. We should not perform measurements during the settle time, and we need to take into account both time intervals for proper synchronization.

If the timing of the SLM is known exactly, we can partially overcome the latency by updating the image on the SLM \emph{before} the current measurement has finished. Once the SLM starts responding (that is, after the latency), the measurement will have completed. With these `pipelined' measurements, enabled by default in OpenWFS, an experiment can be run significantly than without, especially if the settle time is short \cite{vellekoop2008controlling}. 

\begin{figure}
\begin{center}
	\includegraphics[width=0.80\linewidth]{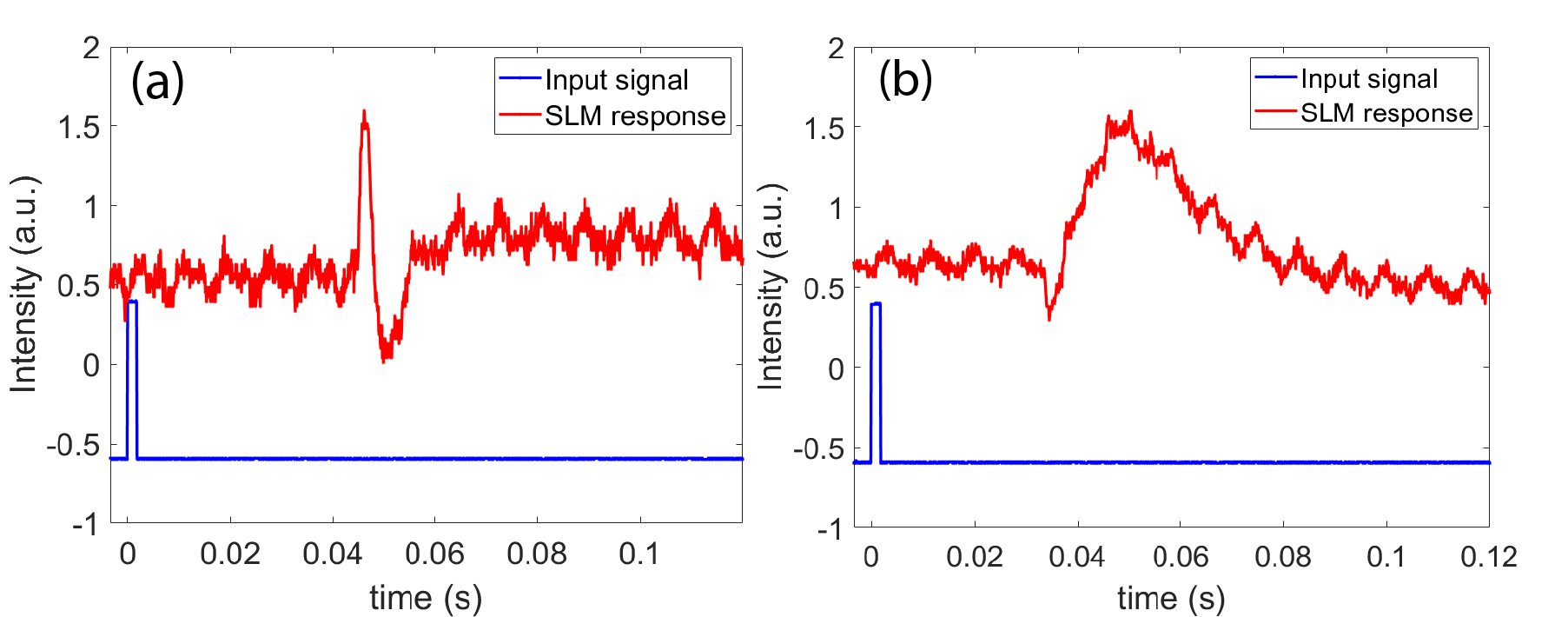}    
\end{center}
\caption{\label{fig:SLM timing} Measured SLM response (a) when the SLM grey value is changed from 0 to 255, and (b) when the grey value is changed from 255 to 0.}
\end{figure}

\subsubsection{Automatic detection and quantification}
To measure idle and settle time interval, the phase of the SLM screen is switched between two phase values while recording the reflected beam from the SLM using a photodiode. Figure~\ref{fig:SLM timing} shows an example of the detected signal measured for Hamamatsu X13138-07 SLM.
The image on this SLM is updated with a refresh rate of 60 frames per second. When we change the grey value from 0 (black screen) to 255 (white screen) the SLM responds as shown in Fig.~\ref{fig:SLM timing}(a) (red curve). The blue pulse represents the input signal whose peak indicates the moment that the new grey value is sent to the SLM. From Fig.~\ref{fig:SLM timing}(a) it can also be estimated that the signal oscillates with a period of \SI{8.3}{ms} (half a frame) representing the phase jitter period. 

After sending the new grey value (white screen) on the SLM, the image on the SLM does not change for at least \SI{33}{ms}, which corresponds to 2 frame periods of latency. After this latency, the SLM takes between 20 and \SI{40}{ms} to stabilize (the \emph{settle time}). It can be seen that the settle time is different for switching from white to black or from black to white.

\subsubsection{Mitigation}
The idle time and settle time should be adjusted to match the measured times (See Fig.~\ref{fig:SLM timing}). Comparing the responses in Fig.~\ref{fig:SLM timing}(a) and Fig.~\ref{fig:SLM timing}(b), we can conclude that the SLM (Hamamatsu X13138-07) responds faster when the grey value is switched from 0 to 255 than when the grey value is changed from 255 to 0. Such information about the SLM timing helps us to design the fast WFS more effectively.
For instance, if we arrange the phase stepping for WFS in ascending order, equivalently changing the grey values from 0 to 255 with a chosen number of steps. It causes the grey value to fall from 255 to 0 in the first iteration of the stepping which takes longer than when the grey value rises from 0 to 255. Therefore, arranging the phase steps in descending order results in avoiding the grey value to fall from 255 to 0, consequently making the WFS faster.

\section{Imperfections in the measurements}
\subsection{Noise in the feedback signal}
\label{sec:noise}
As explained in Section~\ref{sec:phase-stepping}, the WFS measurements consist of modulating part of the SLM segments and finding the required phase essentially by fitting sinusoids. In practice, the recorded feedback signal includes the effect of different types of noise such as read-out noise, shot noise, and laser excess noise. In the presence of noise, the measured feedback signal has deviations from the ideal sinusoid, causing the fidelity to decrease by a factor of \cite{yilmaz2013optimal}
\begin{equation}
\fid{N}=\frac{SNR}{1+SNR}=\frac{P_\text{sig}}{P_\text{sig} + P_\text{noise}}, \label{eq:fidelity_SNR}
\end{equation}
where SNR is the signal-to-noise ratio, and $P_\text{sig}$ and $P_\text{noise}$ are the signal and noise components of the measurement, respectively.

\subsubsection{Automatic detection and quantification}
In the DFT of the phase-stepping experiments (see Section~\ref{sec:phase-stepping}) component $F_0$ refers to the signal offset, and $F_{-1}$ and $F_1$ correspond to the sinusoidal response. All remaining components in the Fourier transform correspond to noise, imperfections in the calibration, or non-linear response from the sample.
Here, we assume that the response is perfectly linear, so that the only effect comes from noise. We furthermore assume white noise, meaning that the noise is distributed equally over all Fourier components. In particular, this means that the total power of the noise in the 1\textsuperscript{st} Fourier component, which also includes the signal, can be estimated as
\begin{equation}
    P_\text{noise}=\frac{1}{P-3}\sum_{\abs{k}>1} \sum_a^N \abs{F_{k,a}}^2,
\end{equation}
where the sum is over all $P$ frequency components of the Fourier transform, excluding the $-1$ and $1$ terms (which include the signal) and the $0$ component (which is the signal offset). Assuming that the noise is independent of the signal, we can estimate the total signal power as
\begin{equation}
    P_\text{sig}=\sum_a^N \abs{F_{1,a}}^2 - P_\text{noise},
\end{equation}
which can be inserted into Eq.~\eqref{eq:fidelity_SNR} to compute the fidelity reduction caused by noise. A further refinement can be done by not treating $F_{1,a}$ as the signal, but to take into account the calibration error and also include all other components $c_k$ in Eq.~\eqref{eq:F-with_calibration}. We do so by constructing a `filtered' signal $\tilde{F}$
\begin{equation}
    \begin{bmatrix}
        \tilde{F}_{k,1}\\
        \tilde{F}_{k,2}\\
        \ldots\\
        \tilde{F}_{k,N}
    \end{bmatrix} = F F^+ \begin{bmatrix}
        F_{k,1}\\
        F_{k,2}\\
        \ldots\\
        F_{k,N}
    \end{bmatrix}
\end{equation}
using the matrix $F$ and its pseudo-inverse $F^+$ as defined in Eq.~\eqref{eq:pseudo-inverse}. We now treat $\tilde{F}$ as the signal instead of $F_{k,1}$.
With a perfect calibration of the SLM, these two approaches are equivalent. With a poorly calibrated SLM, however, this approach allows separating the effect of noise and poor calibration to avoid `double counting' of the calibration fidelity. OpenWFS computes $\fid{N}$ from $\tilde{F}$ and stores the value as \texttt{fidelity\_noise}.

\subsubsection{Mitigation}
One technique to reduce the effect of noise is to increase the size of the area where the feedback signal is collected from. Since more light is collected, the SNR of the measurement increases. However, if we take the feedback area too large, the enhancement will be distributed over multiple targets and decreases as $|\gamma|^2=1/M$, where $M$ is the number of targets \cite{vellekoop2008phase}. 

A different technique to improve the SNR is to perform a pre-optimization with a limited number of optical modes \cite{vellekoop2007focusing,yilmaz2013optimal,tao2017three}. This technique increases the strength of the reference field, thereby increasing the signal strength. The drawback of this technique is that it only works when wavefront shaping for a single target. Many algorithms allow measuring the optimal wavefront for many targets $b$ simulaneously, for example when a camera is used as a detector~\cite{popoff2010measuring}.

It was shown in Ref.~\cite{popoff2010measuring} that modulating multiple segments at a time can be used to greatly increase the SNR of the phase stepping measurements. In these approaches the number of controlled pixels is equal to the number of reference pixels. It leads to an optimally balanced interference of modulated and reference field. This 50/50 choice for the reference and controlled part guarantees the maximal interferometric visibility and thus maximizes $P_\text{sig}$ described above. An example of this technique is to use part of the SLM as reference, and perform measurements of the remaining part of the SLM by grouping the pixels in a Hadamard basis \cite{popoff2010measuring}. Since the fixed reference basically corresponds to non-modulated light, this technique does reduce the fidelity a bit. This reduction can be overcome using a dual-reference technique \cite{mastiani2021noise}, of which a version working in a Fourier basis \cite{mastiani2022wavefront} is available as the \texttt{FourierDualReference} algorithm. 

\subsection{\label{sec:stability}Instability}
A low fidelity may also be the result of an instability in the setup, or a time varying sample. As in any optical setup, there are possible sources of drift, vibrations, and turbulence. The transmission matrix of the sample itself may also vary on very short timescales, such as when the sample is perfused tissue \cite{jang2015relation}.  

\subsubsection{Automatic detection and quantification}
In a transmission geometry, a straightforward way to check the stability of the setup is to capture the transmitted light through the sample before and after the measurements. An exact value for the fidelity cannot easily be determined from these measurements only, since it strongly depends on the nature of the sample. However, it was found that for strongly scattering samples the correlation coefficient between the two images gives a good indication of the reduction of the fidelity due to the instability of the setup \cite{jang2015relation}. 

In OpenWFS, we estimate the effect of decorrelation using the Pearson correlation coefficient,
\begin{equation}
    \fid{D}=\frac{\avg{\left(I_1 - \avg{I_1}_b\right)\left(I_2 - \avg{I_2}_b\right)}_b}{\sqrt{\avg{\left(I_1 - \avg{I_1}_b\right)^2}_b}\sqrt{\avg{\left(I_2 - \avg{I_2}_b\right)^2}_b}},\label{eq:fidelity-decorrelation}
\end{equation}
where $\avg{\cdot}_b$ denotes averaging over all pixels of the output image. For strongly scattering samples, where the intensity has an exponential distribution, this approach is equivalent to that in Ref.~\cite{jang2015relation}. The advantage of Eq.~\eqref{eq:fidelity-decorrelation} is that  $\fid{D}\leq 1$ regardless of the intensity distribution of the transmitted light, and that the effect of any background intensity $\intensityoffset$ is removed. This estimate is available as \texttt{fidelity\_decorrelation} in the result of the \texttt{troubleshoot} function.

\subsubsection{Mitigation}
There are some simple ways to increase the stability of the setup: 1) Covering the setup using boxes around the setup, with heat sources placed outside the boxes, decreases the air turbulence and temperature-induced drift. 2) In the case of using a water immersion objective consider the possibility of bubbles formation and avoid it. 3) Since heating of the SLM can cause drift, send randomly changing patterns to the SLM for some time to `pre-heat' it. 4) In order to define the source of instability, separating the probable source from the rest of the setup can help, for instance, turning off all equipment that may cause vibrations or heating, and comparing the long-term stability with each of these devices turned on individually.

Lastly, it should be noted that one should either perform the full WFS algorithm faster than the persistence time of the dynamic sample \cite{conkey2012high}, or use a dynamic algorithm designed to continuously follow the scattering behaviour of the sample \cite{vellekoop2008phase, dArcoGigan2022PhysicsbasedNeural}.

\subsection{Photobleaching}
\label{sec:photobleaching}
In fluorescence microscopy, another reason that the enhancement can seem lower than it should be, is due to bleaching of the fluorophores. Due to many WFS measurements done on the same volume in the sample, significant photobleaching can occur in the measured spot during the WFS measurements. For this reason, the signal can appear to not have improved all that much, while in reality, the sample simply photobleached. Moreover, a very successful WFS experiment can in some cases improve the focus so much that the sample rapidly photobleaches as soon as the computed phase correction pattern is displayed on the SLM.

\subsubsection{Automatic detection and quantification}
Photobleaching can easily be detected by measuring the same frame before and after the WFS measurements. If the fluorescence signal has significantly decreased due to photobleaching, the frame taken after the measurements will show a lower signal. Note that both frames should be taken with same phase pattern (typically a flat phase pattern) on the SLM, in order to make a fair comparison. To assess the performance of the WFS algorithm, enhancement should be quantified based only on frames frames taken after the WFS measurements.

\subsubsection{Mitigation}
Photobleaching can be reduced by simply reducing the excitation power. Of course, this comes at the cost of a lower signal to noise ratio, meaning that the mitigation strategies presented in Section~\ref{sec:noise} may need to be applied.

\section{Other considerations}
\subsection{Multi-target feedback}
Throughout this paper, we considered the task of optimizing the intensity in a single mode $b$. In practice, however, the feedback often comes from a finite-size region that effectively spans multiple optical modes. In this case, WFS finds a solution that simultaneously maximizes the intensity in multiple modes simultaneously. It can be shown \cite{vellekoop2008phase} that in this case the enhancement is distributed over these modes, effectively causing the total enhancement to reduce with a factor of the number of modes $M$.  

In practice, this means there is a trade-off between the size of the feedback area and the enhancement. Obtaining feedback from a larger area (e.g. by averaging over multiple camera pixels, or using a larger fluorescent guide star) may result in a better SNR, but choosing this area too large causes the enhancement (and also the resolution of the focus) to decrease.

\subsection{Non-linear feedback}
\label{sec:non-linear-feedback}
Feedback in a WFS experiment can come in different forms. An interesting form of feedback is that of multi-photon excitation fluorescence (multi-PEF, or more specifically 2-PEF, 3-PEF, etc.). In multi-PEF, a fluorophore is excited by multiple photons before emitting a fluorescence photon. A 2-photon excitation spectrum of a fluorophore is typically similar to the single-photon excitation spectrum but with the wavelength doubled \cite{xu1996MultiphotonCrosssections}. 
In multi-PEF, the fluorescence signal scales non-linearly with the excitation intensity. If the fluorescence is confined to a small volume (a `guide star'), a narrower spot size will produce a significantly higher non-linear signal. This can be used as feedback signal for WFS \cite{tang2012SuperpenetrationOptical}.

Despite the non-linear characteristic of the feedback, Phase stepping can still be used to retrieve the relative phase with Eq.~\eqref{eq:phase-step-dft}, but ideally $I_p$ follows the shape of a $\cos^{2 n}(\phi_p/2)$ function, with $n$ denoting the order of non-linearity, as explained in Ref.~\cite{sinefeldAdaptiveOpticsMultiphoton2015}. This non-sinusoidality of $I_p$ causes predictable non-zero $F_k$ for ${\abs{k}\geq 2}$, even in the absence of noise and for a perfect SLM response.

Lastly, if the fluorescence signal is not coming from a single point, the exact relationship between the order of the non-linearity and the achievable enhancement is complicated and depends on properties of the excitation source and the structure of the fluorescent material in the sample. \cite{sinefeldAdaptiveOpticsMultiphoton2015, cheng2014MeasurementsMultiphoton, xu1996MultiphotonFluorescence} 
With 2-PEF in a fluorescent dye pool, for example, forming a single focus does not optimize the total fluorescence. Instead, WFS finds a complicated pattern of hot-spots throughout the pool \cite{Katz2014}. If the labelling density is sparse enough, however, iterative WFS converges to a tight focus \cite{Katz2014,osnabrugge2019blind}. 
Lastly, for non-linear feedback, some amount of non-sinusoidality of $I_p$ is to be expected. See also Section~\ref{sec:non-linear-feedback}.

\subsection{Amplitude-only light modulators}
Digital micro-mirror devices (DMDs) are light modulators that modulate only the amplitude of the incident light \cite{conkey2012high,akbulut2011focusing}. DMDs have a much faster switching rate than liquid crystal SLMs, which is an advantage. However, for strongly scattering samples the amplitude-only modulation reduces the fidelity to a maximum of $|\gamma|^2=1/(2\pi)\approx0.16$, \cite{akbulut2011focusing} which is significantly lower than the maximum theoretical fidelity of $\pi/4\approx 0.79$ for phase-only SLMs. For strongly scattering samples, the fidelity can be increased by roughly a factor of $2$ by simply blocking the 0-th order reflection of the DMD \cite{chandrasekaran2014using,vellekoop2012digital}. Alternatively, the DMD can be used in a configuration that enables phase modulation, or even phase-and-amplitude modulation, at the cost of reducing it's effective resolution \cite{conkey2012high, goorden2014}

\section{OpenWFS for automatic troubleshooting\label{sec:openwfs-troubleshooter}}

\definecolor{codegreen}{rgb}{0,0.6,0}
\definecolor{codegray}{rgb}{0.3,0.3,0.3}
\definecolor{codepurple}{rgb}{0.58,0,0.82}
\definecolor{backcolour}{rgb}{0.96,0.96,0.96}

\lstdefinestyle{mystyle}{
    backgroundcolor=\color{backcolour},   
    commentstyle=\color{codegreen},
    keywordstyle=\color{blue},
    numberstyle=\tiny\color{codegray},
    stringstyle=\color{codepurple},
    basicstyle=\ttfamily\footnotesize,
    breakatwhitespace=true,         
    breaklines=true,                 
    captionpos=b,                    
    keepspaces=true,                 
    numbersep=5pt,                  
    showspaces=false,                
    showstringspaces=false,
    showtabs=false,                  
    tabsize=2
}
\lstset{style=mystyle}

\begin{figure}
    \centering
\begin{lstlisting}[language=Python]
import numpy as np
import astropy.units as u
from openwfs.processors import SingleRoi
from openwfs.simulation import SLM, Microscope, Shutter
from openwfs.algorithms import StepwiseSequential, troubleshoot
from openwfs.utilities import set_pixel_size

# === Define virtual devices for a WFS simulation ===
# Note: these virtual-device classes may be replaced
# with real-device classes to run a real WFS experiment

# Define aberration as a pattern of random phases at the pupil plane
aberrations = np.random.uniform(size=(40, 40)) * 2 * np.pi

# Define specimen as an image with several bright pixels
specimen = np.zeros((120, 120))
specimen[60, 60] = 1e4
specimen = set_pixel_size(specimen, pixel_size=200 * u.nm)

# Simulate an SLM with incorrect phase response
# Also simulate a shutter that can turn off the light
# The SLM is conjugated to the back pupil plane
slm = SLM(shape=(100, 100),
          phase_response=(np.arange(256) / 128 * np.pi) * 1.2)
shutter = Shutter(slm.field)

# Simulate a WFS microscope looking at the specimen
sim = Microscope(source=specimen, incident_field=shutter, aberrations=aberrations, wavelength=800 * u.nm)

# Simulate a camera device with gaussian noise and shot noise
cam = sim.get_camera(analog_max=1e4, shot_noise=True, gaussian_noise_std=4.0)

# Define feedback as circular region of interest in the center of the frame
roi_detector = SingleRoi(cam, radius=0.1)

# === Run wavefront shaping experiment ===
# Use the stepwise sequential (SSA) WFS algorithm
alg = StepwiseSequential(feedback=roi_detector, slm=slm, n_x=10, n_y=10, phase_steps=16)

# Define a region of interest to determine background intensity
roi_background = SingleRoi(cam, radius=10)

# Run WFS troubleshooter and output a report to the console
trouble = troubleshoot(algorithm=alg, background_feedback=roi_background,
                       frame_source=cam, shutter=shutter)
trouble.report()
\end{lstlisting}
    \caption{Example code for running a WFS simulation in OpenWFS.\\ (examples/troubleshooter\_demo.py)}
    \label{fig:code-example}
\end{figure}

\begin{figure}
    \centering
\begin{lstlisting}
=== Feedback metrics ===
number of modes (N): 100.000
fidelity_amplitude: 0.726
fidelity_noise: 0.719
fidelity_non_modulated: 0.931
fidelity_phase_calibration: 0.884
expected enhancement: 30.154
measured enhancement: 15.920

=== Frame metrics ===
fidelity_decorrelation: 0.702
average signal offset: 1.63
photobleaching ratio: 1.001
\end{lstlisting}
    \caption{Example output from the \texttt{troubleshoot} function.}
    \label{fig:code-example-output}
\end{figure}

\begin{figure}
    \centering
    \includegraphics[width=0.49\linewidth]{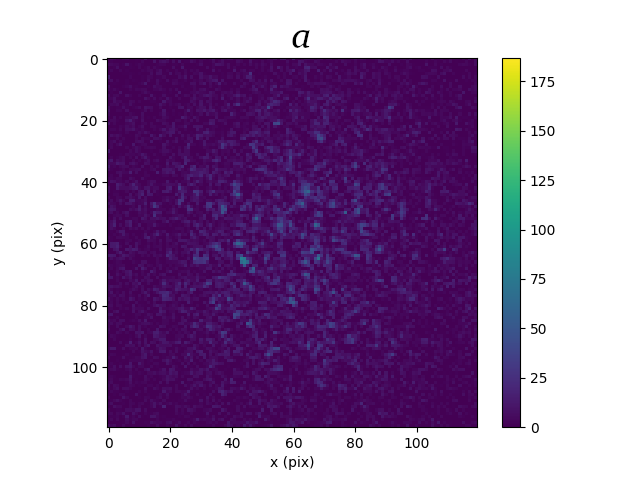}
    \includegraphics[width=0.49\linewidth]{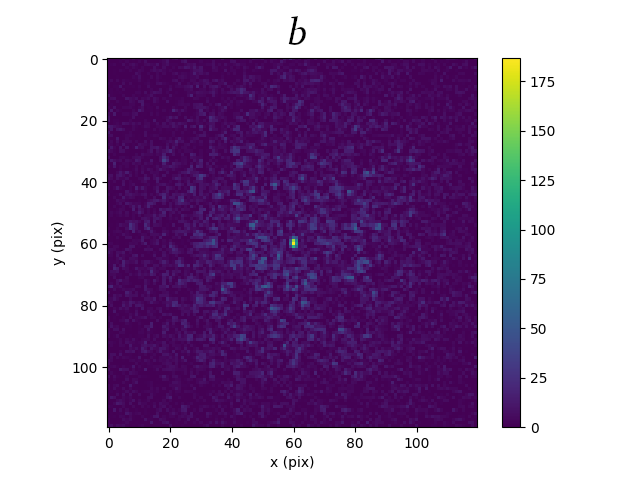}
    \caption{Simulated wavefront shaping with OpenWFS. a. A simulated speckle pattern. b. A focus acquired after simulated wavefront shaping under imperfect circumstances.}
    \label{fig:sim-wfs-with-OpenWFS}
\end{figure}

In this section we show how to run a wavefront shaping experiment with our OpenWFS framework \cite{vellekoop2023OpenwfsGithub}. In the Python script shown in Fig.~\ref{fig:code-example}, we define virtual hardware devices to run a WFS simulation. In this example, we simulate an SLM with an incorrect phase response by defining the SLM object with a custom \texttt{phase\_response} function, and a virtual camera with significant noise. Before running the WFS algorithm, the virtual camera sensor shows a speckle pattern (Fig.~\ref{fig:sim-wfs-with-OpenWFS}a). After running the WFS algorithm, the virtual camera sensor shows a focus (Fig.~\ref{fig:sim-wfs-with-OpenWFS}b). Due to several detrimental effects, the focus is significantly lower than the theoretical maximum. We show that our troubleshooter can detect and approximately quantify the degraded quality of the result and identify the detrimental effects.

The troubleshooter extracts much of its information from the WFS measurements directly. These results are saved as properties of the \texttt{WFSResult} object that is returned by the algorithm.
We can currently simulate the detrimental effects of unequal illumination profile, noise, incorrect phase calibration and non-modulated light. The code is generic such that it can be used both for simulations and for experiments, by replacing the virtual hardware classes with real hardware classes.

We show an example output of the \texttt{troubleshoot} function in Fig.~\ref{fig:code-example-output}. An enhancement of almost 16
has been reached in this experiment. Although this is already a usable enhancement, this is not quite as high as the theoretical value of 100, equal to the number of measured modes. The metrics give insight into what caused the reduced enhancement. The \texttt{fidelity\_amplitude} is significantly lower than 1, indicating that the SLM segments were not equally illuminated. The \texttt{fidelity\_noise} is significantly lower than 1, indicating a significant effect of noise. The \texttt{fidelity\_phase\_calibration} is significantly lower than 1, indicating a problem with the SLM phase calibration. Lastly, all these fidelity estimations are combined to compute an order of magnitude estimate of the expected enhancement. These numbers indicate which are the likely causes of the lower-than-ideal enhancement, and thus help in mitigating their detrimental effects.

\section{Conclusion}
The field of WFS is full of exciting new advances, from interesting new use cases to state-of-the-art algorithms.  However, the most scientific publications only give limited attention to the the technical details of implementing these methods in practice.

In this paper, we described many of the problems that can degrade the performance of a WFS experiment. Anyone starting in this field is likely to encounter one or more of these problems, giving WFS a steep `learning curve' that may frustrate researchers and needlessly slow down progress in this exciting field.

We gave a comprehensive overview of the most common issues that may plague WFS experiments, covering issues caused by imperfections in the feedback measurements, imperfections of the SLM, or simply properties of the modulation scheme. 

We found that many of these issues can be detected and quantified automatically, often using the data that is already available from the WFS experiments themselves. These automatic troubleshooters are integrated in our open source OpenWFS Python package \cite{vellekoop2023OpenwfsGithub}, which also provides all the essential boilerplate code to start working with spatial light modulators, WFS algorithms, and for simulating WFS under realistic experimental conditions. This way, we hope to lower the threshold for starting in this field, and to reduce much of the frustration that may be accompanied with troubleshooting experimental issues.

\section*{Acknowledgments}
This work was supported by Nederlandse Organisatie voor Wetenschappelijk Onderzoek (Vidi grant 14879).

\section*{Data and code availability}
\texttt{openwfs} is available on \url{https://github.com/IvoVellekoop/openwfs} and on the standard PyPI Python package repository. For this manuscript version \texttt{0.1.0rc1} was used.

\bibliographystyle{plain}
\bibliography{main}
\end{document}